\documentstyle[12pt]{article}
\def\vcb{\mid V_{cb} \mid}
\def\vtd{\mid V_{td} \mid}
\def\vub{\mid V_{ub}/V_{cb} \mid}
\def\f{\frac}
\def\o{\over}
\newcommand{\al}{\alpha_s}
\def\kpnn{K^+\rightarrow\pi^+\nu\bar\nu }
\def\kpn{K^+\rightarrow\pi^+\nu\bar\nu}
\def\klpnn{K_L\rightarrow\pi^0\nu\bar\nu}
\def\klpn{K_L\rightarrow\pi^0\nu\bar\nu}

\def\imlt{{\rm Im}\lambda_t}
\def\relt{{\rm Re}\lambda_t}
\def\relc{{\rm Re}\lambda_c}

\newcommand{\be}{\begin{equation}}
\newcommand{\ee}{\end{equation}}
\newcommand{\ra}{\rightarrow}

\newcommand{\Bsg}{$B \ra X_s \gamma$ }

\newcommand{\bea}{\begin{eqnarray}}
\newcommand{\eea}{\end{eqnarray}}
\newcommand{\bd}{\begin{displaymath}}
\newcommand{\ed}{\end{displaymath}}

\newcommand{\kpiee}{$K_L \ra \pi^0 e^+ e^-$ }
\newcommand{\Lms}{\Lambda_{\overline{\rm MS}}}

\begin{document}
\thispagestyle{empty}
\begin{flushright}
 MPI-PhT/95-17 \\
 TUM-T31-85/95 \\
 March 1995
\end{flushright}
\vskip1truecm
\centerline{\Large\bf  Rare Decays, CP Violation and QCD
   \footnote[1]{\noindent Dedicated to the 60th Birthday of Kacper Zalewski.
 To appear in Acta Physica Polonica.}}
  \vskip1truecm
\centerline{\sc Andrzej J. Buras}
\bigskip
\centerline{\sl Technische Universit\"at M\"unchen, Physik Department}
\centerline{\sl D-85748 Garching, Germany}
\vskip0.6truecm
\centerline{\sl Max-Planck-Institut f\"ur Physik}
\centerline{\sl  -- Werner-Heisenberg-Institut --}
\centerline{\sl F\"ohringer Ring 6, D-80805 M\"unchen, Germany}

\vskip1truecm
\centerline{\bf Abstract}
We discuss several aspects of rare decays and CP violation
in the standard model including the impact of the recent top quark
discovery. In particular we review the present status
of next-to-leading QCD calculations in this field stressing their
importance in the determination of the parameters
in the Cabibbo-Kobayashi-Maskawa matrix.
 We emphasize that
the definitive tests of the standard model picture of rare decays and
CP violation will come
through a {\it simultaneous} study of CP asymmetries in $B_{d,s}^0$ decays,
the rare decays
$K^+ \to \pi^+\nu \bar\nu$ and $K_L \to \pi^0\nu\bar\nu$, and
$(B^0_d-\bar B^0_d)/(B^0_s-\bar B^0_s)$.

\newpage
\setcounter{page}{1}

\centerline{\Large\bf  Rare Decays, CP Violation and QCD}
 \vskip1truecm
\centerline{\sc Andrzej J. Buras}
\bigskip
\centerline{\sl Technische Universit\"at M\"unchen, Physik Department}
\centerline{\sl D-85748 Garching, Germany}
\vskip0.6truecm
\centerline{\sl Max-Planck-Institut f\"ur Physik}
\centerline{\sl  -- Werner-Heisenberg-Institut --}
\centerline{\sl F\"ohringer Ring 6, D-80805 M\"unchen, Germany}
\vskip1truecm
\centerline{\bf Abstract}
{\small
We discuss several aspects of rare decays and CP violation
in the standard model including the impact of the recent top quark
discovery. In particular we review the present status
of next-to-leading QCD calculations in this field stressing their
importance in the determination of the parameters
in the Cabibbo-Kobayashi-Maskawa matrix.
 We emphasize that
the definitive tests of the standard model picture of rare decays and
CP violation will come
through a {\it simultaneous} study of CP asymmetries in $B_{d,s}^0$ decays,
the rare decays
$K^+ \to \pi^+\nu \bar\nu$ and $K_L \to \pi^0\nu\bar\nu$, and
$(B^0_d-\bar B^0_d)/(B^0_s-\bar B^0_s)$.
}
\section{Preface}
It is a great privilege and a great pleasure to give this talk
at the symposium celebrating the 60th birthday of Kacper Zalewski.
I have known Kacper during the last 20 years  admiring  him, his research
 and his constructive criticism. I do hope very much to give another
talk on this subject in 2015 at a symposium celebrating Kacper's
80th birthday. I am convinced
that the next 20 years in the field of rare decays and CP violation will
be very exciting and hopefully full of surprises. A 1995 view of this field
is given below.
\section{Setting the Scene}

An important target of particle physics is the determination
 of the unitary $3\times 3$ Cabibbo-Kobayashi-Maskawa
matrix \cite{CAB,KM} which parametrizes the charged current interactions of
 quarks:
\begin{equation}\label{1j}
J^{cc}_{\mu}=(\bar u,\bar c,\bar t)_L\gamma_{\mu}
\left(\begin{array}{ccc}
V_{ud}&V_{us}&V_{ub}\\
V_{cd}&V_{cs}&V_{cb}\\
V_{td}&V_{ts}&V_{tb}
\end{array}\right)
\left(\begin{array}{c}
d \\ s \\ b
\end{array}\right)_L
\end{equation}
The CP violation in the standard model is supposed to arise
from a single phase in this matrix.
It is customary these days to express the CKM-matrix in
terms of four Wolfenstein parameters
\cite{WO} $(\lambda,A,\varrho,\eta)$
with $\lambda=\mid V_{us}\mid=0.22 $ playing the role of an expansion
parameter and $\eta$
representing the CP violating phase:
\begin{equation}\label{2.75}
V_{CKM}=
\left(\begin{array}{ccc}
1-{\lambda^2\over 2}&\lambda&A\lambda^3(\varrho-i\eta)\\ -\lambda&
1-{\lambda^2\over 2}&A\lambda^2\\ A\lambda^3(1-\varrho-i\eta)&-A\lambda^2&
1\end{array}\right)
+O(\lambda^4)
\end{equation}
Because of the
smallness of $\lambda$ and the fact that for each element
the expansion parameter is actually
$\lambda^2$, it is sufficient to keep only the first few terms
in this expansion.

\vspace{6.2cm}
\centerline{Fig. 1}

Following \cite{BLO} one can define the parameters
$(\lambda, A, \varrho, \eta)$ through
\be\label{wop}
s_{12}\equiv\lambda \qquad s_{23}\equiv A \lambda^2 \qquad
s_{13} e^{-i\delta}\equiv A \lambda^3 (\varrho-i \eta)      \ee
where $s_{ij}$ and $\delta$ enter the standard exact
parametrization \cite{PDG}  of the CKM
matrix. This specifies the higher orders terms in (\ref{2.75}).

The definition of $(\lambda,A,\varrho,\eta)$ given in (\ref{wop})
is useful because it allows to improve the accuracy of the
original Wolfenstein parametrization in an elegant manner. In
particular
\begin{equation}\label{CKM1}
V_{us}=\lambda \qquad V_{cb}=A\lambda^2
\end{equation}
\begin{equation}\label{CKM2}
V_{ub}=A\lambda^3(\varrho-i\eta)
\qquad
V_{td}=A\lambda^3(1-\bar\varrho-i\bar\eta)
\end{equation}
where
\begin{equation}\label{3}
\bar\varrho=\varrho (1-\frac{\lambda^2}{2})
\qquad
\bar\eta=\eta (1-\frac{\lambda^2}{2})
\end{equation}
turn out \cite{BLO} to be excellent approximations to the
exact expressions.

A useful geometrical representation of the CKM matrix is the unitarity
triangle obtained by using the unitarity relation
\begin{equation}\label{2.87h}
V_{ud}V_{ub}^* + V_{cd}V_{cb}^* + V_{td}V_{tb}^* =0,
\end{equation}
rescaling it by $\mid V_{cd}V_{cb}^\ast\mid=A \lambda^3$ and depicting
the result in the complex $(\bar\rho,\bar\eta)$ plane as shown
in fig. 1. The lenghts CB, CA and BA are equal respectively to 1,
\begin{equation}\label{2.94a}
R_b \equiv  \sqrt{\bar\varrho^2 +\bar\eta^2}
= (1-\frac{\lambda^2}{2})\frac{1}{\lambda}
\left| \frac{V_{ub}}{V_{cb}} \right|
\qquad
{\rm and}
\qquad
R_t \equiv \sqrt{(1-\bar\varrho)^2 +\bar\eta^2}
=\frac{1}{\lambda} \left| \frac{V_{td}}{V_{cb}} \right|.
\end{equation}

The triangle in fig. 1 is one of the important targets of the contemporary
particle physics. Together with $\mid V_{us}\mid$ and $\mid V_{cb}\mid$
it summarizes the structure of the CKM matrix. In particular the area of
the unrescaled triangle gives a measure of CP violation in the
standard model
\cite{JAR}:
\begin{equation}\label{555}
\mid J_{CP}\mid=
2\cdot ({\rm Area~ of~} \Delta)=
\mid V_{ud}\mid\mid V_{us}\mid\mid V_{ub}\mid\mid V_{cb}\mid \sin\delta=
A^2\lambda^6\bar\eta={\cal O}(10^{-5}).
\end{equation}
This formula shows an important feature of the KM picture of CP
violation: the smallness of CP violation in the standard model is not
necessarily related to the smallness of $\eta$ but to the fact that in this
model the size of CP violating effects is given by products of
small mixing parameters.

\vspace{11.5cm}
\centerline{Fig. 2}

Looking at the expressions for $R_b$ and $R_t$ we also observe that within
the standard model the measurements of four CP
{\it conserving } decays sensitive to $\mid V_{us}\mid$, $\mid V_{ub}\mid$,
$\mid V_{cb}\mid $ and $\mid V_{td}\mid$ can tell us whether CP violation
is predicted in the standard model. This is a very remarkable property of
the Kobayashi-Maskawa picture of CP violation: quark mixing and CP violation
are closely related to each other.

There is of course the very important question whether the KM picture
of CP violation is correct and more generally whether the standard
model offers a correct description of weak decays of hadrons. In order
to answer these important questions it is essential to calculate as
many branching ratios as possible, measure them experimentally and
check if they all can be described by the same set of the parameters
$(\lambda,A,\varrho,\eta)$. In the language of the unitarity triangle
this means that the various curves in the $(\bar\varrho,\bar\eta)$ plane
extracted from different decays should cross each other at a single point
as shown in fig. 2. Moreover the angles $(\alpha,\beta,\gamma)$ in the
resulting triangle should agree with those extracted one day from
CP-asymmetries in B-decays. More about this below.

There is a common belief that during the coming fifteen years we will
certainly witness a dramatic
improvement in the determination of the CKM-parameters analogous
to, although not as precise as, the determination of the parameters
in the gauge boson sector which took place during the recent years.
To this end, however, it is essential not only to perform difficult
experiments but also to have accurate formulae which would allow
a confident and precise extraction of the CKM-parameters from the
existing and future data. We will review what
progress has been done in this direction in Section 4.

Finally it is important to stress that the discovery of the top quark
\cite{CDF,D0} and its mass measurement had an important impact on
the field of rare decays and CP violation reducing considerably one
potential uncertainty. In loop induced K and B
decays the relevant mass parameter is the running current quark mass.
With the pole mass measurement of CDF, $m^{pole}_t=176\pm 13~GeV$,
one has $m_t^*=\bar m_t(m_t)\approx 168\pm 13~GeV$. Similarly the
D0 value $m^{pole}_t=199\pm 30~GeV$ corresponds to
$m_t^*=\bar m_t(m_t)\approx 190\pm 30~GeV$.
 In this review we will
simply denote $m_t^*$ by $m_t$.
\section{Basic Framework}
\subsection{OPE and Renormalization Group}
The basic framework for weak decays of hadrons containing u, d, s, c and
b quarks is the effective field theory relevant for scales $ \mu \ll
M_W, M_Z , m_t$. This framework
brings in local operators which govern ``effectively''
the transitions in question. From the point of view of the decaying
hadrons containing the lightest five quarks this is the only correct
picture we know and also the most efficient one in studying the
presence of QCD. Furthermore it represents the generalization of the Fermi
theory. In this connection it should be mentioned that the usual Feynman
diagram drawings containing full  W-propagators or
$ Z^0-$propagators and top-quark propagators represent
really the happening at scales $ {\cal O}(M_W)$ whereas the true picture
of a decaying hadron is more correctly described by effective vertices
which are represented by local operators in question.

Thus whereas at scales $ {\cal O}(M_W) $ we deal with
the full six-quark theory
containing photon, weak gauge bosons and gluons, at scales ${\cal O}(1 GeV)$
 the
relevant effective theory contains only three light quarks u, d and s,
gluons and the photon. At intermediate energy scales, $\mu={\cal O}(m_b)$
and $\mu={\cal O}(m_c)$,
relevant for b and charm decays effective five-quark
and effective four-quark theories have to be considered respectively.

The usual procedure then is to start at a high energy scale
${\cal O}(M_W) $
and consecutively integrate out the heavy degrees of freedom (heavy with
respect to the relevant scale $ \mu $) from explicitly appearing in the
theory. The word ``explicitly'' is very essential here. The heavy fields
did not disappear. Their effects are merely hidden in the effective
gauge coupling constants, running masses and most importantly in the
coefficients describing the ``effective'' strength of the operators at a
given scale $\mu $, the Wilson coefficient functions.

Operator Product Expansion (OPE) combined with the renormalization group
approach can be regarded as a mathematical formulation of the picture
outlined above.
In this framework the amplitude for a decay $M\to F$ is written as
\begin{equation}\label{OPE}
 A(M \to F) = \frac{G_F}{\sqrt 2} V_{CKM} \sum_i
   C_i (\mu) \langle F \mid Q_i (\mu) \mid M \rangle
\end{equation}
where $M$ stands for the decaying meson, $F$ for a given final state and
$V_{CKM}$ denotes the relevant $CKM$ factor.
$ Q_i(\mu) $ denote
the local operators generated by QCD and electroweak interactions.
$ C_i(\mu) $ stand for the Wilson
coefficient functions (c-numbers).
The scale $ \mu $ separates the physics contributions in the ``short
distance'' contributions (corresponding to scales higher than $\mu $)
contained in $ C_i(\mu) $ and the ``long distance'' contributions
(scales lower than $ \mu $) contained in $ < F \mid Q_i (\mu) \mid M > $.
 By evolving the scale from $ \mu ={\cal O}(M_W) $ down to
lower values of $ \mu $ one transforms the physics information
at scales higher
than $ \mu $ from the hadronic matrix elements into $ C_i(\mu) $. Since no
information is lost this way the full amplitude cannot depend on $ \mu $.
This is the essence of renormalization group equations which govern the
evolution $ (\mu -dependence) $ of $ C_i(\mu) $. This $ \mu $-dependence
must be cancelled by the one present in $\langle  Q_i (\mu)\rangle $.
 It should be
stressed, however, that this cancellation generally involves many
operators due to the operator mixing under renormalization.

The general expression for $ C_i(\mu) $ is given by:
\begin{equation}
 \vec C(\mu) = \hat U(\mu,M_W) \vec C(M_W)
\end{equation}
where $ \vec C $ is a column vector built out of $ C_i $'s.
$\vec C(M_W)$ are the initial conditions which depend on the
short distance physics at high energy scales.
In particular they depend on $m_t$.
$ \hat U(\mu,M_W) $, the evolution matrix,
is given as follows
\begin{equation}\label{UM}
 \hat U(\mu,M_W) = T_g exp \lbrack
   \int_{g(M_W)}^{g(\mu)}{dg' \frac{\hat\gamma^T(g')}{\beta(g')}}\rbrack
\end{equation}
with $g$ denoting QCD effective coupling constant. $ \beta(g) $
governs the evolution of $g$ and $ \hat\gamma $ is the anomalous dimension
matrix of the operators involved. The structure of this equation
makes it clear that the renormalization group approach goes
 beyond the usual perturbation theory.
Indeed $ \hat  U(\mu,M_W) $ sums automatically large logarithms
$ \log M_W/\mu $ which appear for $ \mu<<M_W $. In the so called leading
logarithmic approximation (LO) terms $ (g^2\log M_W/\mu)^n $ are summed.
The next-to-leading logarithmic correction (NLO) to this result involves
summation of terms $ (g^2)^n (\log M_W/\mu)^{n-1} $ and so on.
This hierarchic structure gives the renormalization group improved
perturbation theory.

As an example let us consider only QCD effects and the case of a single
operator. Keeping the first two terms in the expansions of
 $\gamma(g)$ and $\beta(g)$ in powers of $g$:
\begin{equation}
\gamma (g) = \gamma^{(0)} \frac{\alpha_{QCD}}{4\pi} + \gamma^{(1)}
\frac{\alpha^2_{QCD}}{16\pi^2}
\quad , \quad
 \beta (g) = - \beta_0 \frac{g^3}{16\pi^2} - \beta_1
\frac{g^5}{(16\pi^2)^2}
\end{equation}
and inserting these expansions into (\ref{UM}) gives:
\begin{equation}\label{UMNLO}
 U (\mu, M_W) = \Biggl\lbrack 1 + {{\alpha_{QCD} (\mu)}\over{4\pi}} J
\Biggl\rbrack \Biggl\lbrack {{\alpha_{QCD} (M_W)}\over{\alpha_{QCD} (\mu)}}
\Biggl\rbrack^P \Biggl\lbrack 1 - {{\alpha_{QCD} (M_W)}\over{4\pi}} J
\Biggl\rbrack
\end{equation}
where
\begin{equation}
P = {{\gamma^{(0)}}\over{2\beta_0}},~~~~~~~~~ J = {{P}\over{\beta_0}}
\beta_1 - {{\gamma^{(1)}}\over{2\beta_0}}.
\end{equation}
General
formulae for $ \hat U (\mu, M_W) $ in the case of operator mixing and
valid also for electroweak effects can be found in ref.\cite{BJLW}.
The leading
logarithmic approximation corresponds to setting $ J = 0 $ in (\ref{UMNLO}).
\subsection{Classification of Operators}
 Below we give six classes of operators which play the
dominant role in the phenomenology of weak decays. Typical diagrams in
the full theory from which these operators originate are indicated
 and shown in Fig. 3 . The cross in Fig. 3d indicates
that magnetic penguins originate from the mass-term on the external
line in the usual QCD or QED penguin diagrams.
The six classes are given as follows:

{\bf Current--Current (Fig. 3a):}
\begin{equation}
Q_1 = (\bar s_{\alpha} u_{\beta})_{V-A}\;(\bar u_{\beta} d_{\alpha})_{V-A}
{}~~~~~~Q_2 = (\bar s u)_{V-A}\;(\bar u d)_{V-A}
\end{equation}

{\bf QCD--Penguins (Fig. 3b):}
\begin{equation}
Q_3 = (\bar s d)_{V-A}\sum_{q=u,d,s}(\bar qq)_{V-A}~~~~~~
 Q_4 = (\bar s_{\alpha} d_{\beta})_{V-A}\sum_{q=u,d,s}(\bar q_{\beta}
       q_{\alpha})_{V-A}
\end{equation}
\begin{equation}
 Q_5 = (\bar sd)_{V-A} \sum_{q=u,d,s}(\bar qq)_{V+A}~~~~~
 Q_6 = (\bar s_{\alpha} d_{\beta})_{V-A}\sum_{q=u,d,s}
       (\bar q_{\beta} q_{\alpha})_{V+A}
\end{equation}

{\bf Electroweak--Penguins (Fig. 3c):}
\begin{equation}
Q_7 = {3\over 2}\;(\bar s d)_{V-A}\sum_{q=u,d,s}e_q\;(\bar qq)_{V+A}
{}~~~~~ Q_8 = {3\over2}\;(\bar s_{\alpha} d_{\beta})_{V-A}\sum_{q=u,d,s}e_q
        (\bar q_{\beta} q_{\alpha})_{V+A}
\end{equation}
\begin{equation}
 Q_9 = {3\over 2}\;(\bar s d)_{V-A}\sum_{q=u,d,s}e_q(\bar q q)_{V-A}
{}~~~~~Q_{10} ={3\over 2}\;(\bar s_{\alpha} d_{\beta})_{V-A}\sum_{q=u,d,s}e_q\;
       (\bar q_{\beta}q_{\alpha})_{V-A}
\end{equation}

{\bf Magnetic--Penguins (Fig. 3d):}
\begin{equation}
Q_{7\gamma}  =  \f{e}{8\pi^2} m_b \bar{s}_\alpha \sigma^{\mu\nu}
          (1+\gamma_5) b_\alpha F_{\mu\nu}\qquad
Q_{8G}     =  \f{g}{8\pi^2} m_b \bar{s}_\alpha \sigma^{\mu\nu}
   (1+\gamma_5)T^a_{\alpha\beta} b_\beta G^a_{\mu\nu}
\end{equation}

{\bf $\Delta S = 2 $ and $ \Delta B = 2 $ Operators (Fig. 3e):}
\begin{equation}
Q(\Delta S = 2)  = (\bar s d)_{V-A} (\bar s d)_{V-A}~~~~~
 Q(\Delta B = 2)  = (\bar b d)_{V-A} (\bar b d)_{V-A}
\end{equation}

{\bf Semi--Leptonic Operators (Fig. 3f):}
\begin{equation}\label{9V}
Q_{9V}  = (\bar b s )_{V-A} (\bar e e)_{V}~~~~~
Q_{10A}  = (\bar b s )_{V-A} (\bar e e)_{A}
\end{equation}
\begin{equation}
Q(\nu\bar\nu)  = (\bar s d)_{V-A} (\nu\bar\nu)_{V-A}~~~~~
Q(\mu\bar\mu)  = (\bar s d)_{V-A} (\mu\bar\mu)_{V-A}
\end{equation}

\vspace{16.6cm}
\centerline{Fig. 3}

\subsection{Towards Phenomenology}
The rather formal expression for the decay amplitudes given in
(\ref{OPE}) can always be cast in the form \cite{PBE}:
\begin{equation}\label{PBEE}
A(M\to F)=\sum_i B_i V_{CKM}^{i} \eta^{i}_{QCD} F_i(m_t,m_c)
\end{equation}
which is more useful for phenomenology. In writing (\ref{PBEE})
we have generalized (\ref{OPE}) to include several CKM factors.
$F_i(m_t,m_c)$, the Inami-Lim functions,
 result from the evaluation of loop diagrams with
internal top and charm exchanges (see fig. 3) and may also depend
solely on $m_t$ or $m_c$. In the case of current-current operators
$F_i$ are mass independent. The factors $\eta^{i}_{QCD}$ summarize
the QCD corrections which can be calculated by formal methods
discussed above. Finally $B_i$ stand for nonperturbative factors
related to the hadronic matrix elements of the contributing
operators: the main theoretical uncertainty in the whole enterprise.
In semi-leptonic decays such as $K\to \pi\nu\bar\nu$,
the non-perturbative $B$-factors can fortunately be determined from
leading tree level decays such as $K^+\to \pi^0 e^+\nu$ reducing
or removing the non-perurbative uncertainty. In non-leptonic
decays this is generally not possible and we have to rely on
existing non-perturbative methods. A well known example of a
$B_i$-factor is the renormalization group invariant parameter
$B_K$ \cite{BSS} defined by
\begin{equation}\label{bk}
B_K=B_K(\mu)\left[\alpha_s(\mu)\right]^{-2/9}
\qquad
\langle \bar K^{o}\mid Q(\Delta S=2)\mid K^{o}\rangle=
\frac{8}{3} B_K(\mu)F_K^2 m_K^2
\end{equation}
$B_K$ plays an important role in the phenomenology of CP violation
in $K \to \pi\pi$. We will encounter several examples of
 (\ref{PBEE}) below.

\section{Weak Decays Beyond Leading Logarithms}
\subsection{General Remarks}
Until 1989 most of the calculations in the field of weak
decays were done in the leading logarithmic approximation.
An exception was the important work of Altarelli et al.\cite{ALTA}
who in 1981 calculated NLO QCD corrections to the Wilson
coefficients of the current-current operators.

Today the effective hamiltonians for weak decays are
available at the next-to-leading level for the most important
and interesting cases due to a series of publications devoted
to this enterprise beginning with the work of Peter Weisz and myself
in 1989 \cite{BW}. The list of the existing calculations is given in
table 1. We will discuss this list briefly below. A detailed review
of the existing NLO calculations will appear soon \cite{BBL}.

Let us recall why NLO calculations are important for the
phenomenology of weak decays.

\begin{itemize}
\item The NLO is first of all necessary to test the validity of
the renormalization group improved perturbation theory.
\item Without going to NLO the QCD scale $\Lambda_{\overline{MS}}$
extracted from various high energy processes cannot be used
meaningfully in weak decays.
\item Due to renormalization group invariance the physical
amplitudes do not depend on the scales $\mu$ present in $\alpha_s$
or in the running quark masses, in particular $m_t(\mu)$,
$m_b(\mu)$ and $m_c(\mu)$. However
in perturbation theory this property is broken through the truncation
of the perturbative series. Consequently one finds sizable scale
ambiguities in the leading order, which can be reduced considerably
by going to NLO.
\item In several cases the central issue of the top quark mass dependence
is strictly a NLO effect.
\end{itemize}

\begin{table}
\begin{center}
\begin{tabular}{|l|l|}
\hline
\bf \phantom{XXXXXX} Decay & \bf \phantom{XX} Reference~~~ \\
\hline
\hline
\multicolumn{2}{|c|}{$\Delta F=1$ Decays} \\
\hline
current-current operators     & \cite{ALTA,BW} \\
QCD penguin operators         & \cite{BJLW1,BJLW,ROMA1,ROMA2} \\
electroweak penguin operators & \cite{BJLW2,BJLW,ROMA1,ROMA2} \\
magnetic penguin operators    & \cite{MisMu:94}  \\
$Br(B)_{SL}$                  & \cite{ALTA,Buch:93,Bagan} \\
\hline
\multicolumn{2}{|c|}{Particle-Antiparticle Mixing} \\
\hline
$\eta_1$                   & \cite{HNa} \\
$\eta_2,~\eta_B$           & \cite{BJW} \\
$\eta_3$                   & \cite{HNb} \\
\hline
\multicolumn{2}{|c|}{Rare K- and B-Meson Decays} \\
\hline
$K^0_L \rightarrow \pi^0\nu\bar{\nu}$, $B \rightarrow l^+l^-$,
$B \rightarrow X_{\rm s}\nu\bar{\nu}$ & \cite{BB1,BB2} \\
$K^+   \rightarrow \pi^+\nu\bar{\nu}$, $K_L \rightarrow \mu^+\mu^-$
                                      & \cite{BB3} \\
$K^+\to\pi^+\mu\bar\mu$               & \cite{BB5} \\
$K_L \rightarrow \pi^0e^+e^-$         & \cite{BLMM} \\
$B\rightarrow X_s e^+e^-$           & \cite{Mis:94,BuMu:94} \\
\hline
\end{tabular}
\end{center}
\centerline{}
\caption{References to NLO Calculations}
\end{table}

\subsection{Current-Current Operators}
The NLO corrections to the coefficients of $Q_1$ and $Q_2$ have been
first calculated by Altarelli et al.\cite{ALTA}
 using the Dimension Reduction
Scheme (DRED) for $\gamma_5$. In 1989 these coefficients have been
calculated in DRED, NDR and HV schemes for $\gamma_5$ by Peter Weisz and
myself \cite{BW}. The result for DRED obtained by the Italian group
has been
confirmed. The coefficients $C_1$ and $C_2$ show a rather strong
renormalization scheme dependence which in physical quantities
should be cancelled by the one present in the matrix elements of
$Q_1$ and $Q_2$. This cancellation has been shown explicitly in
\cite{BW} demonstrating thereby the compatibilty of the results for
$C_1$ and $C_2$ in DRED, NDR and HV schemes. A recent discussion
of $C_1(\mu)$ and $C_2(\mu)$ in these schemes can be found in
\cite{AB:95c}.
\subsection{NLO Corrections to $B_{SL}$}
A direct physical application of the NLO corrections to $C_1$ and
$C_2$ discussed above is the calculation of the non-leptonic width
for B-Mesons which is relevant for the theoretical prediction of
the inclusive semileptonic branching ratio:
\begin{equation}
B_{SL}=\frac{\Gamma(B\to Xe\nu)}{\Gamma_{SL}(B)+\Gamma_{NL}(B)}
\end{equation}
This calculation can be done within the spectator model corrected
for small non-perturbative corrections \cite{Bigi}
 and more important gluon
bremsstrahlung and virtual gluon corrections. The latter cancell the
scheme and $\mu$ dependences of $C_i(\mu)$. The calculation of $B_{SL}$
for massless final quarks has been done by Altarelli et al.\cite{ALTA}
 in the
DRED scheme and by Buchalla \cite{Buch:93} in the HV scheme.
 The results of these papers agree with each other.

It is well known that the inclusion of QCD corrections in the
spectator model, lowers $B_{SL}$ which otherwise
would be roughly $16\%$. Unfortunately the theoretical branching ratio
based on the QCD calculation of refs. \cite{ALTA,Buch:93}  give typically
$B_{SL}=12.5-13.5\%$ \cite{AP:92}
whereas the experimental world average \cite{PDG} is
\begin{equation}\label{BEXP}
B^{exp}_{SL}=(10.43\pm 0.24)\%
\end{equation}
The inclusion of the leading non-perturbative correction
${\cal O}(1/m_b^2)$ lowers slightly the theoretical
prediction but gives only $\Delta_{NP} B_{SL}=-0.2\%$ \cite{Bigi}.
On the other hand mass effects in the QCD corrections to $B_{SL}$
seem to play an important role.
Bagan et al. \cite{Bagan} using partially the results of Hokim and Pham
\cite{Pham} have demonstrated that the inclusion
of mass effects in
the QCD calculations of refs.\cite{ALTA,Buch:93} (in particular in the
decay $b \to c\bar c s$ (see also \cite{Voloshin} )) and taking into account
various renormalization
scale uncertainties improves the situation considerably.
Bagan et al. find \cite{Bagan}:
\begin{equation}
B_{SL}=(12.0\pm 1.4)\% \quad {\rm and} \quad \bar B_{SL}=(11.2\pm1.7)\%
\end{equation}
for the pole quark masses and $\overline{MS}$ masses respectively.
Within existing uncertainties, this result does not disagree
significantly with the experimental value, although it is still
somewhat on the high side.
\subsection{$\Delta S=2$ and $\Delta B=2 $ Transitions}
The $M_{12}$ amplitude describing the $K^{0}-\bar K^{0}$ mixing is
given as follows
\begin{equation}
M_{12}(\Delta S=2)=\frac{G_F^2}{12\pi^2}F_K^2 B_K m_K M_W^2
\left[\lambda_c^{*2}\eta_1 S(x_c) +\lambda_t^{*2}\eta_2 S(x_t)
+2\lambda_c^{*}\lambda_t^{*}\eta_3 S(x_c,x_t)\right]
\end{equation}
with $x_i=m_i^2/M_W^2$, $\lambda_i=V_{id}V_{is}^{*}$,
 $S(x_i)$ denoting the Inami-Lim functions
resulting from box diagrams and $\eta_i$ representing QCD corrections.
The parameter $B_K$ is defined in (\ref{bk}).
The corresponding amplitude for the $B_d^{o}-\bar B_d^{o}$ mixing is
dominated by the box diagrams with top quark exchanges and given by
\begin{equation}
\mid M_{12}(\Delta B=2)\mid =\frac{G_F^2}{12\pi^2}F_B^2 B_B m_B M_W^2
\mid V_{td} \mid^2 \eta_B S(x_t)
\end{equation}
where we have set $V_{tb}=1$. A similar formula exists for
$B_s^{o}-\bar B_s^{o}$.
For  $m_t<<M_W$ and in the leading order
$\eta_i$ have been calculated by Gilman and Wise \cite{GIL}.
Generalization to $m_t={\cal O}(M_W)$ gives roughly
\cite{BBH,KSY,Flynn,DFP}
\begin{equation}\label{GW}
\eta_1=0.85\qquad \eta_2=0.62\qquad \eta_3=0.36\qquad\eta_B=0.60
\end{equation}

As of 1995 the coefficients $\eta_i$ and $\eta_B$ are known including
NLO corrections. The coefficients $\eta_2$ and $\eta_B$ have been
calculated in \cite{BJW} and $\eta_1$ and $\eta_3$ in \cite{HNa} and
\cite{HNb} respectively.
It has been stressed in these papers that the LO results for $\eta_i$
in (\ref{GW}) suffer from sizable scale uncertainties, as large
as $\pm 20\%$ for $\eta_1$ and $\pm 10\% $ for the remaining $\eta_i$.
As demonstrated in \cite{BJW,HNa,HNb} these uncertainties
are considerably reduced in the products like
$\eta_1 S(x_c),~\eta_2 S(x_t),~\eta_3 S(x_c,x_t)$ and $\eta_B S(x_t)$
  provided NLO corrections
are taken into account. For $m_c=\bar m_c(m_c)=1.3\pm0.1~GeV$ and
$m_t=\bar m_t(m_t)=170\pm 15~GeV$
one finds:
\begin{equation}\label{KNLO}
\eta_1=1.3\pm 0.2\qquad \eta_2=0.57\pm0.01\qquad \eta_3=0.**\pm0.04
\qquad
\eta_B=0.55\pm0.01
\end{equation}
where the "**" in $\eta_3$ will be public soon \cite{HNb}.
It should be stressed that $\eta_i$ given here are so defined that
the relevant $B_K$ and $B_B$ non-perturbative factors (see (\ref{bk}))
are renormalization group invariant.

Let us list the main implications of these results:
\begin{itemize}
\item
The enhancement of $\eta_1$ implies the enhacement of the short
distance contribution to the $K_L-K_S$ mass difference so that for
$B_K=3/4$ as much as $80\%$ of the experimental value can be
attributed to this contribution \cite{HNa}.
\item
The improved calculations of $\eta_2$ and $\eta_3$
combined with the analysis of the CP violating parameter $\varepsilon_K$
allow an improved determination of the parameters $\eta$ and $\varrho$
in the CKM matrix \cite{BLO,HNb}.
\item
Similarly the improved calculation of $\eta_B$ combined with the
analysis of $B^0_d-\bar B^0_d$ mixing allows an improved determination
of the element $\mid V_{td}\mid$ \cite{BLO}:
\begin{equation}
\mid V_{td} \mid=
8.7\cdot 10^{-3}\left [
\frac{200~MeV}{\sqrt{B_B}F_B}\right ]
\left [\frac{170~GeV}{\bar m_t(m_t)} \right ]^{0.76}
\left [\frac{x_d}{0.72} \right ]^{0.5}
\left [\frac{1.50~ps}{\tau_B} \right ]^{0.5}
\end{equation}
This using all uncertainties (see below) gives:
\begin{equation}\label{vtd}
\mid V_{td} \mid = (9.6\pm 3.0)\cdot 10^{-3}
\quad
=>\quad (9.3 \pm 2.5 )\cdot 10^{-3}
\end{equation}
with the last number obtained after the inclusion of the
$\varepsilon$-analysis \cite{BLO}.
\end{itemize}
Concerning the parameter $B_K$, the most recent analyses
using the lattice methods
\cite{SH0,Ishizuka} ($B_K=0.83\pm 0.03$) and the $1/N$ approach
 of \cite{BBG0}
modified somewhat in \cite{Bijnens} give results in the ball park
of the $1/N$ result $B_K=0.70\pm 0.10$ obtained long
time ago \cite{BBG0}. In particular the analysis of Bijnens and Prades
\cite{Bijnens} seems to have explained the difference between these values
for $B_K$ and the lower values obtained using the QCD Hadronic Duality
approach \cite{Prades} ($B_K=0.39\pm 0.10$) or using SU(3) symmetry and
 PCAC
($B_K=1/3$) \cite{Donoghue}. This is gratifying because such low values for
$B_K$ would require $m_t>250~GeV$ in order to explain the experimental
value of $\varepsilon$ \cite{AB,BLO,HNb}.

There is a vast literature on the lattice calculations of $F_B$. The
most recent results are somewhat lower than quoted a few years ago.
Based on a review by Chris Sachrajda \cite{Chris}, the recent extensive
study by Duncan et al. \cite{Duncan} and the analyses in \cite{Latt}
we conclude:
$F_{B_d}=(180\pm40)~MeV$. This together with the earlier result of
the European Collaboration for $B_B$, gives
$F_{B_d}\sqrt{B_{B_d}}=195\pm 45~MeV$.
The reduction of the error in this important quantity is desirable.
These results for $F_B$ are compatible with the results obtained using
QCD sum rules (eg.\cite{QCDS}). An interesting upper bound
$F_{B_d}<195~MeV$ using QCD dispersion relations has also recently
been obtained \cite{BGL}.
\subsection{$\Delta S=1$ Hamiltonian and $\varepsilon'/\varepsilon$}
The effective Hamiltonian for $\Delta S=1$ transitions is given
as follows:
\begin{equation}\label{dels1}
{\cal H}_{eff}(\Delta S=1) = \f{G_F}{\sqrt{2}} V_{us}^* V_{ud}
 \sum_{i=1}^{10} \left[ z_i(\mu)+\tau y_i(\mu)\right] Q_i
\end{equation}
where $\tau=-(V_{td}V_{ts}^*)/(V_{ud}V_{us}^*)$.
The coefficients of all ten operators are known
including NLO QCD and QED effects in NDR and HV schemes due to
the independent work of Munich and Rome groups
\cite{BJLW1,BJLW2,BJLW,ROMA1,ROMA2}.
 The results of
both groups agree with each other. A direct application of these
results is the calculation of
Re($\varepsilon'/\varepsilon$) which measures the ratio of direct
to indirect
CP violation in $K\to\pi\pi$ decays. In the standard model
$\varepsilon'/\varepsilon $ is governed by QCD penguins and
electroweak (EW)
penguins \cite{GIL0}. In spite of being suppressed by $\alpha/\alpha_s$
relative to QCD penguin contributions, the electroweak penguin contributions
have to be included because of the additional enhancement factor
$ReA_0/ReA_2=22$ relative to QCD penguins. Moreover with increasing $m_t$
the EW-penguins become increasingly important \cite{FLYNN,BBH} and entering
$\varepsilon'/\varepsilon$ with the opposite sign to QCD-penguins suppress
this ratio for large $m_t$. For $m_t\approx 200~GeV$ the ratio can even
be zero \cite{BBH}.
This strong cancellations between these two contributions was one of
the prime motivations for the NLO calculations performed in Munich
and Rome. Although these calculations can be regarded as an important
step towards a reliable theoretical prediction for
$\varepsilon'/\varepsilon$ the situation is clearly not satisfactory
at present.
Indeed $\varepsilon'/\varepsilon$ is plagued with uncertainties related to
non-perturbative B-factors which multiply $m_t$ dependent functions in a
formula like (\ref{PBEE}). Several of these B-factors can be extracted from
leading
CP-conserving $K\to\pi\pi$ decays \cite{BJLW}. Two important B-factors
($B_6=$ the dominant QCD penguin ($Q_6$) and $B_8=$ the dominant electroweak
 penguin ($Q_8$))
cannot be determined this way and one has to use lattice or $1/N$ methods
to predict Re($\varepsilon'/\varepsilon$).

 An analytic formula for
Re($\varepsilon'/\varepsilon$) as a function of
$m_t,~\Lambda_{\overline{MS}},~B_6,~B_8,~m_s$ and $V_{CKM}$ can be found
in \cite{BLAU}. A very simplified version of this formula is given as follows
\begin{equation}\label{7e}
{\rm Re}(\frac{\varepsilon'}{\varepsilon})=12\cdot 10^{-4}\left [
\frac{\eta\lambda^5 A^2}{1.7\cdot 10^{-4}}\right ]
\left [\frac{150~MeV}{\bar m_s(m_c)} \right ]^2
\left [\frac{\Lms^{(4)}}{300~MeV} \right ]^{0.8}
[B_6-Z(x_t)B_8]
\end{equation}
where $Z(x_t)$ is given in (\ref{11k}). Note the strong dependence on
$\Lms$ pointed out in \cite{BJLW}.
For $m_t=170\pm13~GeV$ and $\bar m_s(m_c)\approx 150\pm20~MeV$ \cite{Jamin}
and using $\varepsilon_K$-analysis to determine
$\eta$ one finds using the formulae in \cite{BJLW,BLAU} roughly
\begin{equation}\label{8}
1\cdot 10^{-4} \leq Re(\frac{\varepsilon'}{\varepsilon})\leq 15\cdot 10^{-4}
\end{equation}
if $B_6=1.0\pm 0.2$ and $B_8=1.0\pm 0.2$ are used.
Such values are found in the $1/N$
approach \cite{BBG} and using lattice methods: \cite{SH1}
and \cite{SH1,SH2} for $B_6$ and $B_8$ respectively.
A very recent analysis of the Rome group \cite{ROMA3} gives a smaller
range,
 $Re(\varepsilon'/\varepsilon)=(3.1\pm 2.5)\cdot 10^{-4}$, which is
 however compatible
with (\ref{8}). Similar results are found with hadronic matrix elements
calculated in the chiral quark model \cite{Stefano}.
However $\varepsilon'/\varepsilon$ obtained in \cite{DORT} is
substantially larger and about $2 \cdot 10^{-3}$.

The experimental situation on Re($\varepsilon'/\varepsilon$) is unclear
at present.
 While
the result of NA31 collaboration at CERN with Re$(\varepsilon'/\varepsilon)
= (23 \pm 7)\cdot 10^{-4}$ \cite{WAGNER} clearly indicates
direct CP violation, the value of E731 at Fermilab,
Re$(\varepsilon'/\varepsilon) = (7.4 \pm 5.9)\cdot 10^{-4}$
\cite{GIBBONS} is compatible with superweak theories \cite{WO1} in which
$\varepsilon'/\varepsilon = 0$.
The E731 result is in the ball park of the theoretical estimates.
The NA31 value appears a bit high compared to the range given in
(\ref{8}).

 Hopefully, in about five years the
experimental situation concerning $\varepsilon'/\varepsilon$ will be
clarified through the improved measurements by the two collaborations
at the $10^{-4}$ level and by experiments at the $\Phi$ factory in
Frascati.
One should also hope that the theoretical situation of
$\varepsilon'/\varepsilon$ will improve by then to confront the new data.
\subsection{$\Delta B=1$ Effective Hamiltonian}
The effective hamiltonian for $\Delta B=1$ transitions involving
operators $Q_1,..Q_{10}$ (with corresponding changes of flavours)
is also known including NLO corrections \cite{BJLW}. It has been
used in the study of CP asymmetries in B-decays \cite{Fleischer:94}.
\subsection{$K\to\pi^o e^+e^-$}
The effective Hamiltonian for $K\to\pi^0 e^+e^-$  is given
as follows:
\begin{equation}\label{dels}
{\cal H}_{eff}(K\to\pi^0 e^+e^-) = \f{G_F}{\sqrt{2}} V_{us}^* V_{ud}
 \left[\sum_{i=1}^{6,9V} \left[ z_i(\mu)+\tau y_i(\mu)\right] Q_i
+\tau y_{10A}(M_W)Q_{10A}\right]
\end{equation}
where $Q_{9V}$ and $Q_{10A}$ are given by (\ref{9V})
 with $\bar b s$ replaced by $\bar s d $.

 Whereas in $K \to \pi \pi$ decays the CP violating
contribution is a tiny part of the full amplitude and the direct CP
violation is expected to be at least by three orders of magnitude
smaller than the indirect CP violation, the corresponding hierarchies
are very different for the rare decay $K_L\to\pi^o e^+e^-$ .
At lowest order in
electroweak interactions (single photon, single Z-boson or double
W-boson exchange), this decay takes place only if CP symmetry is
violated \cite{GIL1}.
Moreover, the direct CP violating contribution is predicted to be
larger than
the indirect one. The CP conserving contribution to the amplitude
comes from a two photon exchange, which
although higher order in $\alpha$ could in principle be sizable.
  The studies in
 \cite{Seghal,PICH} indicate however that the
CP conserving part is  smaller than the direct CP
violating contribution.

The size of the indirect CP violating contribution will be
known once the CP conserving decay $K_S \to \pi^0 e^+ e^-$ has been
measured \cite{BARR}. On the other hand the direct CP violating
contribution can
be fully calculated as a function of $m_t$, CKM parameters and the
QCD coupling constant $\alpha_s$. There are practically no theoretical
uncertainties related to hadronic matrix elements in this part,
because the relevant matrix elements of the operators $Q_{9V}$ and
$Q_{10A}$ can be extracted from the well-measured decay
$K^+\to \pi^0 e^+ \nu$.

Restricting the attention to the CP violating parts of the coefficients
$C_{9V}$ and $C_{10V}$ and factoring out the relevant CKM factor as well
as $\alpha/2\pi$ one finds \cite{BLMM}
\begin{equation}\label{pbe7v}
\tilde y_{9V}  =  P_0 +\frac{Y(x_t)}{ \sin^2 \theta_W}-
4  Z(x_t) \qquad,\qquad
\tilde y_{10A} =  - \f  {Y(x_t)}{\sin^2 \theta_W}.
\end{equation}
where, to a very good approximation for $140~GeV\leq m_t \leq 230~GeV$,
\begin{equation}\label{11k}
Y(x_t) = 0.315\cdot x_t^{0.78}, \hspace{2cm} Z(x_t) = 0.175\cdot x_t^{0.93}.
\end{equation}
The next-to-leading QCD corrections to $P_0$
have been calculated in \cite{BLMM}
reducing certain ambiguities present in
leading order analyses \cite{GIL2} and enhancing the leading order value
typically from $P_0(LO)=1.9$ to $P_0(NLO)=3.0$
The final result for the branching ratio is given by
\begin{equation}\label{9a}
Br(K_L \to \pi^0 e^+ e^-)_{dir} = 6.3\cdot 10^{-6}(Im\lambda_t)^2
(\tilde y_{7A}^2 + \tilde y_{7V}^2)
\end{equation}
where
$Im \lambda_t = Im (V_{td} V^*_{ts})$.
For $m_t=170\pm 10~GeV$ one finds \cite{BLMM}
\begin{equation}\label{12}
Br(K_L\to\pi^0 e^+ e^-)_{dir}=(5.\pm 2.)\cdot 10^{-12}
\end{equation}
where the error comes dominantly from the uncertainties in the CKM
parameters.
This should be compared with the present estimates of the other two
contributions:  $Br(K_L\to\pi^o e^+e^-)_{indir}\leq 1.6\cdot 10^{-12}$
and $Br(K_L\to\pi^o e^+e^-)_{cons}\approx(0.3-1.8)\cdot 10^{-12}$ for
the indirect
CP violating and the CP conserving contributions respectively \cite{PICH}.
Thus direct
CP violation is expected to dominate this decay.

The present experimental
bounds
\begin{equation}
Br(K_L\to\pi^0 e^+ e^-) \leq\left\{ \begin{array}{ll}
4.3 \cdot 10^{-9} & \cite{harris} \\
5.5 \cdot 10^{-9} & \cite{ohl} \end{array} \right.
\end{equation}
are still by three orders of magnitude away from the theoretical
expectations in the Standard Model. Yet the prospects of getting the
required sensitivity of order $10^{-11}$--$10^{-12}$ in five years are
encouraging \cite{CPRARE}.

\subsection {$B\to X_s\gamma$}
The effective hamiltonian for $B\to X_s\gamma$ at scales $\mu=O(m_b)$
is given by
\be \label{Heff_at_mu}
{\cal H}_{eff}(b\to s\gamma) = - \f{G_F}{\sqrt{2}} V_{ts}^* V_{tb}
\left[ \sum_{i=1}^6 C_i(\mu) Q_i + C_{7\gamma}(\mu) Q_{7\gamma}
+C_{8G}(\mu) Q_{8G} \right]
\ee
where in view of $|V_{us}^*V_{ub} / V_{ts}^* V_{tb}| < 0.02$
 we have neglected the term proportional to $V_{us}^*V_{ub}$.

 The perturbative QCD effects are very important in this decay.
They are known
\cite{Bert,Desh} to enhance $B\to X_s\gamma$ in
the SM by 2--3
times, depending on the top quark mass. Since the first analyses
in \cite{Bert,Desh} a lot of progress has been made in calculating
the QCD effects begining with the work in \cite{Grin,Odon}. We will
briefly summarize this progress.

A peculiar feature of the renormalization group analysis
in $B\to X_s\gamma$ is that the mixing under infinite renormalization
between
the set $(Q_1...Q_6)$ and the operators $(Q_{7\gamma},Q_{8G})$ vanishes
at the one-loop level. Consequently in order to calculate
the coefficients
$C_{7\gamma}(\mu)$ and $C_{8G}(\mu)$ in the leading logarithmic
approximation, two-loop calculations of ${\cal{O}}(e g^2_s)$
and ${\cal{O}}(g^3_s)$
are necessary. The corresponding NLO analysis requires the evaluation
of the mixing in question at the three-loop level.

At present, the coefficients $C_{7\gamma}$ and $C_{8G}$ are only known
in the leading logarithmic approximation.
However the peculiar feature of this decay mentioned above caused
that the first fully correct calculation of the leading  anomalous
dimension matrix has been obtained only in 1993 \cite{CFMRS:93,CFRS:94}.
It has been
confirmed subsequently in \cite{CCRV:94a,CCRV:94b,Mis:94}.
In order to extend these calculations beyond the leading order
one would have to calculate $\hat\gamma^{(1)}_s$ and $O(\alpha_s)$
corrections to the initial conditions $\vec C(M_W)$.
The $6\times 6$ two-loop
submatrix of $\hat\gamma^{(1)}_s$ involving the operators
$Q_1.....Q_6$ is the same as in section 4.5. The two-loop mixing
in the sector $(Q_{7\gamma},Q_{8G})$ has been calculated only
last year \cite{MisMu:94}.
The three loop mixing between
the set $(Q_1...Q_6)$ and the operators $(Q_{7\gamma},Q_{8G})$
 has not be done. The $O(\alpha_s)$
corrections to $C_{7\gamma}(M_W)$ and $C_{8G}(M_W)$ have been considered
in \cite{Yao1}. Gluon corrections to the matrix elements of magnetic
penguin operators have been calculated in \cite{AG1,AG2}.

The leading
logarithmic calculations
\cite{Grin,CFRS:94,CCRV:94a,Mis:94,AG1,BMMP:94}
   can be summarized in a compact form,
as follows:
\be\label{main}
\f{Br(B \ra X_s \gamma)}{Br(B \ra X_c e \bar{\nu}_e)}
 =  \f{|V_{ts}^* V_{tb}|^2}{|V_{cb}|^2}
\f{6 \alpha_{QED}}{\pi f(z)} |C^{(0)eff}_{7\gamma}(\mu)|^2
\ee
 where
$C^{(0)eff}_{7\gamma}(\mu)$ is the effective coefficient
for which an analytic expression can be found in \cite{BMMP:94},
 $z = {m_c}/{m_b}$, and
$f(z)$ is the phase space factor in the semileptonic
b-decay.
The expression given above is based on the
spectator model corrected for short-distance QCD effects.
Support for this approximation
comes from the $1/m_b $ expansions.
Indeed the spectator
model has been shown to correspond to the leading order approximation
in the $1/m_b$ expansion.
The next corrections appear at the ${\cal O}(1/m_b^2)$
level. The latter terms have been studied by several authors
\cite{Chay,Bj,Bigi} with the result that they affect Br(\Bsg) and
Br($B \ra X_c e \bar{\nu}_e$) by only a few percent.

A critical analysis of theoretical and
experimental
uncertainties present in the prediction for Br(\Bsg) based on the
formula (\ref{main}) has been made in \cite{BMMP:94} giving
\be
Br(B \ra X_s\gamma)_{TH} = (2.8 \pm 0.8) \times 10^{-4}.
\label{theo}
\ee
where the error is dominated by the uncertainty in
choice of the renormalization scale
$m_b/2<\mu<2 m_b$ as first stressed by Ali and Greub \cite{AG1} and confirmed
in \cite{BMMP:94}.
	Since \Bsg is dominated by QCD effects, it is not surprising
that this scale-uncertainty in the leading order
is particularly large.

The \Bsg decay has already been measured and as such appears to be the only
unquestionable signal of penguin contributions! In 1993
CLEO reported \cite{CLEO}
$Br(B \ra K^* \gamma) = (4.5 \pm 1.5 \pm 0.9) \times 10^{-5}.$
In 1994 first measurement of the inclusive rate has been
presented by CLEO \cite{CLEO2}:
\be
Br(B \ra X_s\gamma) = (2.32 \pm 0.57 \pm 0.35) \times 10^{-4}.
\label{incl}
\ee
where the first error is statistical and the second is systematic.
This result agrees with (\ref{theo}) very well although
the theoretical and experimental errors should be decreased in
the future in order to reach a definite conclusion and to see
whether some contributions beyond the standard model
 such as present in the
Two-Higgs-Doublet Model (2HDM)
or in the Minimal Supersymmetric Standard
Model (MSSM) are required. In any case the agreement of the
theory with data is consistent with the large QCD enhancement
of \Bsg. Without this enhancement the theoretical prediction
would be at least by a factor of 2 below the data.

	Fig. 4  presents the SM prediction for the inclusive
\Bsg branching ratio including the uncertainties discussed in
\cite{BMMP:94} together with the CLEO results represented by the
shaded regions.
We stress that the theoretical result (the error bars) has been
obtained prior
to the experimental result. Since the theoretical error
is dominated by scale ambiguities  a  complete
NLO analysis is very desirable.
Such a complete next-to-leading
calculation of \Bsg is described in \cite{BMMP:94} in general terms.
As demonstrated formally there
 the cancellation of the dominant $\mu$-dependence in the leading
order can be achieved by calculating the relevant two-loop
matrix element of
the dominant four-quark operator $Q_2$.
 This matrix element is however renormalization-scheme
dependent and moreover mixing with other operators takes place.
This scheme dependence can only be cancelled by calculating
$\hat{\gamma}^{(1)}$ in the same renormalization scheme.
This however requires the three loop mixing mentioned above.

\vspace{11.1cm}
\centerline{Fig. 4}

In this connections we would like to comment on an analysis
of \cite{Ciu:94} in which the known two-loop mixing in the sector
$(Q_1....Q_6)$ (see table 1) has been added to the leading order
 analysis of \Bsg.
Strong renormalization scheme dependence of the resulting
branching ratio has been found, giving the branching ratio
$(1.7\pm 0.2)\cdot 10^{-4}$ and $(2.3 \pm0.3)\cdot 10^{-4}$
at $\mu=5~GeV$ for HV and NDR
schemes respectively. It has also been observed that whereas
in the HV scheme the $\mu$ dependence has been weakened,
it remained still strong in the NDR scheme. In our opinion
the partial cancellation of the $\mu$-dependence in the HV
scheme is rather accidental and has nothing to do with the
cancellation of the $\mu$-dependence discussed above. The latter
requires the evaluation of finite parts in two-loop matrix
elements of the four-quark operators $(Q_1.......Q_6)$.
On the other hand the strong scheme dependence in the partial
NLO analysis presented in \cite{Ciu:94} demonstrates very
clearly the need for a full analysis.
In view of this discussion we think
that the decrease of the branching ratio for \Bsg
relative to the LO prediction, found
in \cite{Ciu:94} and given by
$Br(B\to s\gamma)=(1.9\pm 0.2\pm 0.5)\cdot 10^{-4}$,
is still premature and one
should wait until the full NLO analysis has been done.

\subsection{$B\to X_s e^+e^-$ Beyond Leading Logarithms}
The effective hamiltonian for $B\to X_s e^+e^-$ at scales $\mu=O(m_b)$
is given by
\be \label{Heff2_at_mu}
{\cal H}_{eff}(b\to s e^+e^-) =
{\cal H}_{eff}(b\to s\gamma)  - \f{G_F}{\sqrt{2}} V_{ts}^* V_{tb}
\left[ C_{9V}(\mu) Q_{9V}+
C_{10A}(M_W) Q_{10A}    \right]
\ee
where again we have neglected the term proportional to $V_{us}^*V_{ub}$
and ${\cal H}_{eff}(b\to s\gamma)$ is given in (\ref{Heff_at_mu}).
In addition to the operators relevant for $B\to X_s\gamma$,
there are two new operators
$Q_{9V}$ and $Q_{10A}$
which appeared already in the decay \kpiee
except for an appropriate change of quark flavours
and the fact that now $\mu={\cal O}(m_b)$ instead of
$\mu={\cal O}(1~GeV)$ should be considered. There is a large literature
on this dacay. In particular Hou et al \cite{HWS:87} stressed
the strong dependence of $B\to X_s e^+e^-$ on $m_t$.
Further references to phenomenology can be found in \cite{BuMu:94}.
Here we concentrate on QCD corrections.

The special feature of $C_{9V}(\mu)$ compared to the coefficients
of the remaining operators contributing to $B\to X_s e^+e^-$ is the
large logarithm represented by $1/\al$ in $P_0$ in a formula
like (\ref{pbe7v}).
 Consequently the renormalization group improved
perturbation theory for $C_{9V}$ has the structure $ {\cal O}(1/\al) +
{\cal O}(1) + {\cal O}(\al)+ \ldots$ whereas the corresponding series
for the remaining coefficients is $ {\cal O}(1) + {\cal O}(\al)+
\ldots$. Therefore in order to find the next-to-leading ${\cal O}(1)$
term in the branching ratio for $B\to X_s e^+e^-$,
 the full two-loop renormalization group analysis
 has to be performed in order to find $C_{9V}$, but the
coefficients of the remaining operators should be taken in the leading
logarithmic approximation. This is gratifying because, as we
discussed above, the coefficients
of the magnetic operators $Q_{7\gamma}$ and $Q_{8G}$
are known only in the leading logarithmic approximation.

The coefficient $C_{9V}(\mu)$ has been calculated
 over the last years with increasing precision by several
groups \cite{GSW:89,GDSN:89,CRV:91,Mis:94} culminating in two complete
next-to-leading QCD calculations
\cite{Mis:94,BuMu:94} which agree with each other.
In particular in \cite{BuMu:94} the coefficient $C_{9V}$ has been
calculated in NDR and HV schemes. Calculating the matrix elements
of the operators $(Q_1,....Q_6)$ in the spectator model the
scheme independence of the resulting physical amplitude has been
demonstrated.

An extensive numerical analysis of the differential decay rate
including NLO corrections has been presented in \cite{BuMu:94}.
As an example we show in fig. 5 the differential decay rate $R(\hat s)$
divided by $\Gamma(B \to X_c e \bar\nu)$ as
a function of $\hat s=(p_{e^+}+p_{e^-})^2/m_b^2$ for $m_t=170~GeV$ and
$\Lambda_{\overline{MS}}=225~MeV$. We observe that the QCD suppression
in the leading order \cite{GSW:89} is substantially weakened by the
inclusion of NLO corrections. Similar result has been obtained by
Misiak \cite{Mis:94}. The $1/m^2_b$ corrections calculated in \cite{FALK}
enhance these results by roughly $10\%$.

\vspace{10.98cm}
\centerline{Fig. 5}

\subsection{$K_L\to\pi^o\nu\bar\nu$, $K^+\to\pi^+\nu\bar\nu$,
$ K_L\to\mu\bar\mu$, $B\to\mu\bar\mu$ and $B\to X_s\nu\bar\nu$}
$K_L\to\pi^o\nu\bar\nu$ and $K^+\to\pi^+\nu\bar\nu$ are the theoretically
cleanest decays in the field of rare K-decays. Similarly
$B\to\mu\bar\mu$ and $B\to X_s\nu\bar\nu$ are the theoretically
cleanest decays in the field of rare B-decays.
$K_L\to\pi^o\nu\bar\nu$,
$B\to\mu\bar\mu$ and $B\to X_s\nu\bar\nu$
 are dominated by short distance loop diagrams
involving the top quark.  $K^+\to\pi^+\nu\bar\nu$ receives
additional sizable contributions from internal charm exchanges.
The decay $K_L\to \mu\bar\mu$ receives substantial long distance
contributions and  consequently suffers from large theoretical
uncertainties. This is very unfortunate because this is the only
rare Kaon decay which has already been measured.
The most accurate is the measurement from Brookhaven \cite{PRINZ}:
\begin{equation}\label{princ}
Br(K_L\to \bar\mu\mu) = (6.86\pm0.37)\cdot 10^{-9}
\end{equation}
which is somewhat lower than the KEK-137 result:
$(7.9\pm 0.6 \pm 0.3)\cdot 10^{-9}$ \cite{Akagi}.
For the short distance contribution I find using the formulae of
\cite{BB3}:
\begin{equation}
Br(K_L\to \bar\mu\mu)_{SD} = (1.5\pm 0.8)\cdot 10^{-9}
\end{equation}
Details on this decay can be found in \cite{PRINZ,BB3}.
More promising from theoretical point of view is the parity-violating
asymmetry in $K^+\to \pi^+\mu^+\mu^-$ \cite{GENG,BB5}.

The next-to-leading QCD corrections to all these decays
have been calculated in a series of papers by Buchalla and
myself \cite{BB1,BB2,BB3,BB5}. These calculations
considerably reduced the theoretical uncertainties
due to the choice of the renormalization scales present in the
leading order expressions \cite{DDG}. Since the relevant hadronic matrix
elements of the weak currents entering $K\to \pi\nu\bar\nu$
can be measured in the leading
decay $K^+ \rightarrow \pi^0 e^+ \nu$, the resulting theoretical
expressions for Br( $K_L\to\pi^o\nu\bar\nu$) and Br($K^+\to\pi^+\nu\bar\nu$)
  are
only functions of the CKM parameters, the QCD scale
 $\Lambda \overline{_{MS}}$
 and the
quark masses $m_t$ and $m_c$.
The long distance contributions to
$K^+ \rightarrow \pi^+ \nu \bar{\nu}$ have been
considered in \cite{RS} and found to be very small: two to three
orders of magnitude smaller than the short distance contribution
at the level of the branching ratio.
The long distance contributions to $K_L\to\pi^o\nu\bar\nu$ are negligible
as well. Similar comments apply to $B\to\mu\bar\mu$ and
$B\to X_s\nu\bar\nu$ except that $B\to\mu\bar\mu$ depends on the
B-meson decay constant $F_B$ which brings in the main theoretical
uncertainty.

The explicit expressions for $Br(\kpnn)$ and $Br(\klpnn)$ are given as
 follows
\be\label{bkpn}
Br(\kpn)=4.64\cdot 10^{-11}
\cdot\left[\left({\imlt\o\lambda^5}X(x_t)\right)^2+
\left({\relc\o\lambda}P_0(K^+)+{\relt\o\lambda^5}X(x_t)\right)^2
\right]            \ee
\be\label{bklpn}
Br(K_L\to\pi^0\nu\bar\nu)=1.94\cdot 10^{-10}
\cdot\left({\imlt\o\lambda^5}X(x_t)  \right)^2
\end{equation}
Here
\begin{equation}
\imlt=\eta A^2\lambda^5 \qquad
\relt=-(1-\frac{\lambda^2}{2})A^2\lambda^5(1-\bar\varrho)
\end{equation}
and $\relc=-\lambda (1-\lambda^2/2)$.
$X(x_t)$ is given to an excellent accuracy by
\begin{equation}\label{xt}
X(x_t) = 0.65\cdot x_t^{0.575}
\end{equation}
where the NLO correction calculated in \cite{BB2} is included if
$m_t\equiv\bar m_t(m_t)$.
 Next $P_0(K^+)=0.40\pm0.09$ \cite{BB3,BB4} is a function of $m_c$ and
 $\Lambda_{\overline{MS}}$ and includes the residual uncertainty
due to the renormalization scale $\mu$. The absence of $P_0$ in
(\ref{bklpn}) makes $\klpnn$ theoretically even cleaner than $\kpnn$.

Similarly for $B_s\to \mu\bar\mu$ one has \cite{BB2}
\begin{equation}
Br(B_s\to \mu\bar\mu)=
4.1\cdot 10^{-9}\left [ \frac{F_{B_s}}{230~MeV}\right ]^2
\left [\frac{\bar m_t(m_t)}{170~GeV} \right ]^{3.12}
\left [\frac{\mid V_{ts}\mid}{0.040} \right ]^2
\left [\frac{\tau_{B_s}}{1.6 ps} \right ]
\end{equation}

The impact of NLO calculations is best illustrated by giving the
scale uncertainties in the leading order and after the inclusion
of the next-to-leading corrections:
\begin{equation}
Br(\kpn)=(1.00\pm0.20)\cdot 10^{-10}\quad =>\quad
(1.00\pm0.05)\cdot 10^{-10}
\end{equation}
\begin{equation}
Br(\klpnn)=(3.00\pm0.30)\cdot 10^{-11}\quad =>\quad
(3.00\pm0.04)\cdot 10^{-11}
\end{equation}
\begin{equation}
Br(B_s\to \mu\bar\mu)=(4.10\pm0.50)\cdot 10^{-9}\quad =>\quad
(4.10\pm0.05)\cdot 10^{-9}
\end{equation}
The reduction of the scale uncertainties is truly impressive.

The present experimental bound on $Br(K^+\to \pi^+\nu\bar\nu)$
is $5.2 \cdot 10^{-9}$ \cite{Atiya}. An improvement by one order
of magnitude is expected at AGS in Brookhaven for the coming years.
The present upper bound on $Br(K_L\to \pi^0\nu\bar\nu)$ from
Fermilab experiment E731 is $10^{-5}$. FNAL-E799 expects to reach
the accuracy ${\cal O}(10^{-8})$ and the future experiments at FNAL
and KEK will hopefully be able to reach the standard model
expectations. The latter are given for both decays at present as follows:
\begin{equation}
Br(\kpn)=(1.1\pm 0.4)\cdot 10^{-10}\quad,\quad
Br(\klpnn)=(3.0\pm 2.0)\cdot 10^{-11}
\end{equation}
\section{Finalists}
\subsection{General Remarks}
{}From tree level K decays sensitive to $V_{us}$ and tree level B decays
 sensitive to $V_{cb}$ and $V_{ub}$ we have \cite{PDG}:
\begin{equation}\label{2}
\lambda=0.2205\pm0.0018
\qquad
\mid V_{cb} \mid=0.040\pm0.004\quad =>\quad A=0.83\pm 0.08
\end{equation}
\begin{equation}\label{2.94}
\left| \frac{V_{ub}}{V_{cb}} \right|=0.08\pm0.03
\quad => \quad
\sqrt{\varrho^2+\eta^2} =0.36\pm0.14
\end{equation}
where the error on $\mid V_{cb}\mid $ still remains a subject of intensive
discussions \cite{Bigi,VCB}.

A large part in the errors quoted in
(\ref{2}) and (\ref{2.94}) results from theoretical (hadronic)
 uncertainties discussed by Ikarus Bigi and Thomas Mannel at this symposium.
Consequently even if the data from CLEO II improves in the
future, it is difficult to imagine at present that
in the tree level B-decays
a better accuracy than $\Delta\vcb=\pm 2\cdot 10^{-3}$ and
$\Delta\vub=\pm 0.01$ ($\Delta R_b=\pm 0.04$) could be achieved unless some
dramatic improvements in the theory will take place.

The question then arises whether it is possible at all to determine the
CKM parameters without any hadronic uncertainties.
As demonstrated in \cite{AB94} this is indeed possible although it will
require heroic experimental efforts.
To this end one has to go to the loop induced decays or transitions
which are fully governed by short distance
physics and study simultaneously CP asymmetries in B-decays.
 In this manner clean and
precise determinations of $\vcb$, $\vub$, $\vtd$, $\varrho$ and $\eta$
can be achieved. Since the relevant measurements will take place only
in the next decade, what follows is really a 21st century story.

It is known that many loop induced decays contain also hadronic uncertainties
\cite{AB94A} related to long distance contributions or poorly known $B_i$
factors.
Examples are $B^0-\bar B^0$ mixing, $\varepsilon_K$ and
 $\varepsilon'/\varepsilon$ discussed above.
Let us in this connection recall the expectations from a "standard" analysis
of the unitarity triangle ( see figs. 1 and 2 )
which is based on $\varepsilon_K$, $x_d$
giving the size of $B^0-\bar B^0$ mixing,
$\vcb$ and $\vub$ with the last two extracted from tree level decays.
As a typical analysis \cite{BLO} shows, even with optimistic assumptions
about the theoretical and experimental errors it will be difficult to
achieve the accuracy better than $\Delta\varrho=\pm 0.15$ and
$\Delta\eta=\pm 0.05$ this way.
More promising at least from the
theoretical point of view are the following four:
\begin{itemize}
\item
CP-Asymmetries in $B^o$-Decays
\item
$K_L\to\pi^o\nu\bar\nu$
\item
$\kpnn$
\item
$(B^o_d-\bar B_d^o)/(B^o_s-\bar B_s^o)$
\end{itemize}
Let us summarize their main virtues one-by-one.
\subsection{CP-Asymmetries in $B^o$-Decays}
The CP-asymmetry in the decay $B_d^\circ \rightarrow \psi K_S$ allows
 in the standard model
a direct measurement of the angle $\beta$ in the unitarity triangle
without any theoretical uncertainties. This has been first pointed out
by Bigi and Sanda \cite {BSANDA}, analyzed in detail already in \cite{BSS}
and during the past years
discussed by many authors \cite{NQ}.
 Similarly the decay
$B_d^\circ \rightarrow \pi^+ \pi^-$ gives the angle $\alpha$, although
 in this case strategies involving
other channels are necessary in order to remove hadronic
uncertainties related to penguin contributions
\cite{CPASYM}.
The determination of the angle~$\gamma$ from CP asymmetries in neutral
B-decays is more difficult but not impossible
\cite{RF}. Also charged B decays could be useful in
this respect \cite{Wyler}.
We have for instance
\begin{equation}\label{113c}
 A_{CP}(\psi K_S)=-\sin(2\beta) \frac{x_d}{1+x_d^2},
\qquad
   A_{CP}(\pi^+\pi^-)=-\sin(2\alpha) \frac{x_d}{1+x_d^2}
\end{equation}
where we have neglected QCD penguins in $A_{CP}(\pi^+\pi^-)$.
Since in the usual unitarity triangle  one side is known,
it suffices to measure
two angles to determine the triangle completely. This means that
the measurements of $\sin 2\alpha$ and $\sin 2\beta$ can determine
the parameters $\varrho$ and $\eta$.
The main virtues of this determination are as follows:
\begin{itemize}
\item No hadronic or $\Lambda_{\overline{MS}}$ uncertainties.
\item No dependence on $m_t$ and $V_{cb}$ (or A).
\end{itemize}
As various analyses \cite{BLO,ALI,ROMA3} of the unitarity triangle
show, $\sin(2\beta)$ is expected to be large:
$\sin(2\beta)\approx 0.6\pm 0.2$.
The predictions for $\sin(2\gamma)$ and $\sin(2\alpha)$ are very
uncertain on the other hand.

\subsection{$K_L\to\pi^o\nu\bar\nu$}
As we have discussed above $K_L\to\pi^o\nu\bar\nu$ is the theoretically
cleanest decay in the field of rare K-decays.
Moreover it proceeds almost entirely through direct CP violation \cite{LI}.
The next-to-leading QCD calculation \cite{BB2}
 reduced the theoretical uncertainty
due to the choice of the renormalization scales present in the
leading order expression
 to $\pm 1\%$.  Since the long distance contributions to
 $K_L\to\pi^o\nu\bar\nu$ are negligible,
 the resulting theoretical
expression for Br( $K_L\to\pi^o\nu\bar\nu$)
given by (see (\ref{bklpn}) and (\ref{xt}))
\begin{equation}
Br(\klpnn)=1.50 \cdot 10^{-5}\eta^2\mid V_{cb}\mid^4 x_t^{1.15}
\end{equation}
 is
only a function of the CKM parameters and  $m_t$.
 The main features of this decay are:
\begin{itemize}
\item No hadronic uncertainties
\item  $\Lambda_{\overline{MS}}$ and renormalization scale uncertainties
at most $\pm 1\%$ \cite{BB2}.
\item Strong dependence on $m_t$ and $V_{cb}$ (or A).
\end{itemize}
\subsection{$\kpnn$}
$K^+\to\pi^+\nu\bar\nu$ is CP conserving and receives
contributions from
both internal top and charm exchanges.
The NLO corrections \cite{BB3} to this decay reduced
the theoretical uncertainty
due to the choice of the renormalization scales present in the
leading order expression
to $\pm 5\%$.
$K^+\to\pi^+\nu\bar\nu$ is then the second best decay in the field
of rare decays. Compared to $\klpnn$ it receives additional uncertainties
due to $m_c$ and the related renormalization scale. Also its QCD scale
dependence is stronger.
The main features of this decay are:
\begin{itemize}
\item Hadronic uncertainties below $1\%$ \cite{RS}
\item  $\Lambda_{\overline{MS}}$, $m_c$ and renormalization scales
 uncertainties at most $\pm (5-10)\%$ \cite{BB3}.
\item Strong dependence on $m_t$ and $V_{cb}$ (or A).
\end{itemize}
\subsection{$(B^o_d-\bar B_d^o)/(B^o_s-\bar B_s^o)$ }
Measurement of $B^o_d-\bar B^o_d$ mixing parametrized by $x_d$ together
with  $B^o_s-\bar B^o_s$ mixing parametrized by $x_s$ allows to
determine $R_t$:
\begin{equation}\label{107b}
R_t = \frac{1}{\sqrt{R_{ds}}} \sqrt{\frac{x_d}{x_s}} \frac{1}{\lambda}
\qquad
R_{ds} = \frac{\tau_{B_d}}{\tau_{B_s}} \cdot \frac{m_{B_d}}{m_{B_s}}
\left[ \frac{F_{B_d} \sqrt{B_{B_d}}}{F_{B_s} \sqrt{B_{B_s}}} \right]^2
\end{equation}
where $R_{ds}$ summarizes SU(3)--flavour breaking effects.
Note that $m_t$ and $V_{cb}$ dependences have been eliminated this way
 and $R_{ds}$
contains much smaller theoretical
uncertainties than the hadronic matrix elements in $x_d$ and $x_s$
separately.
Provided $x_d/x_s$ has been accurately measured a determination
of $R_t$ within $\pm 10\%$ should be possible. Indeed the most
recent lattice result \cite{Duncan} gives $F_{B_d}/F_{B_s}=1.22\pm0.04$.
It would be useful to know $B_{B_s}/B_{B_d}$ with similar precision.
For $B_{B_s}=B_{B_d}$ I find $R_{ds}=0.62\pm 0.07$. Consequently
rescaling the results of \cite{BLO}, obtained for $R_{ds}=1$, the
range $12 < x_s < 39 $ follows. Such a large mixing will not be
easy to measure. The main features of $x_d/x_s$ are:
\begin{itemize}
\item No $\Lambda_{\overline{MS}}$, $m_t$ and $V_{cb}$ dependence.
\item Hadronic uncertainty in SU(3)--flavour breaking effects of
      roughly $\pm 10\%$.
\end{itemize}
Because of the last feature, $x_d/x_s$ cannot fully compete in the clean
determination of CKM parameters with CP asymmetries in B-decays and
with $\klpnn$. Although $\kpnn$ has smaller hadronic uncertainties
than $x_d/x_s$, its dependence on $\Lambda_{\overline{MS}}$ and  $m_c$
puts it in the same class as $x_d/x_s$ \cite{AB94A}.
\subsection{$\sin (2\beta)$ from $K\to \pi\nu\bar\nu$}
It has been pointed out in \cite{BH} that
measurements of $Br(\kpn)$ and $Br(\klpn)$ could determine the
unitarity triangle completely provided $m_t$ and $V_{cb}$ are known.
In view of the strong dependence of these branching ratios on
$m_t$ and $V_{cb}$ this determination is not precise however \cite{BB4}.
On the other hand it has been noticed \cite{BB4} that the $m_t$ and
$V_{cb}$ dependences drop out in the evaluation of $\sin(2\beta)$.
Introducing the "reduced" branching ratios
\begin{equation}\label{19}
B_+={Br(\kpn)\o 4.64\cdot 10^{-11}}\qquad
B_L={Br(\klpn)\o 1.94\cdot 10^{-10}}
\end{equation}
one finds
\begin{equation}\label{22}
\sin (2\beta)=\frac{2 r_s(B_+, B_L)}{1+r^2_s(B_+, B_L)}
\end{equation}
where
\begin{equation}\label{21}
r_s(B_+, B_L)=
{\sqrt{(B_+-B_L)}-P_0(K^+)\o\sqrt{B_L}}
\end{equation}
so that $\sin (2\beta)$
does not depend on $m_t$ and $V_{cb}$.
Here $P_0(K^+)=0.40\pm0.09$ \cite{BB3,BB4} is a function of $m_c$ and
 $\Lambda_{\overline{MS}}$ and includes the residual uncertainty
due to the renormalization scale $\mu$.
Consequently $\kpnn$ and $\klpnn$ offer a
clean
determination of $\sin (2\beta)$ which can be confronted with
the one possible in $B^0\to\psi K_S$ discussed above.
Any difference in these two determinations
would signal new physics.
Choosing
$Br(\kpn)=(1.0\pm 0.1)\cdot 10^{-10}$ and
$Br(\klpn)=(2.5\pm 0.25)\cdot 10^{-11}$,
one finds \cite{BB4}
\begin{equation}\label{26}
\sin(2 \beta)=0.60\pm 0.06 \pm 0.03 \pm 0.02
\end{equation}
where the first error is "experimental", the second represents the
uncertainty in $m_c$ and  $\Lambda_{\overline{MS}}$ and the last
is due to the residual renormalization scale uncertainties. This
determination of $\sin(2\beta)$ is competitive with the one
expected at the B-factories at the beginning of the next decade.
\subsection{Precise Determinations of the CKM Matrix}
Using the first two finalists and $\lambda=0.2205\pm 0.0018$
\cite{LR}
it is possible to determine all the parameters
of the CKM matrix without any hadronic uncertainties
\cite{AB94}.
With $a\equiv \sin(2\alpha)$, $b\equiv \sin(2\beta)$
and $Br(K_L)\equiv Br(\klpnn)$
one determines $\varrho$, $\eta$
and $\vcb$ as follows \cite{AB94}:
\begin{equation}\label{5}
\bar\varrho = 1-\bar\eta r_{+}(b)\quad ,\quad
\bar\eta=\frac{r_{-}(a)+r_{+}(b)}{1+r_{+}^2(b)}
\end{equation}
\begin{equation}\label{9b}
\mid V_{cb}\mid=0.039\sqrt\frac{0.39}{\eta}
\left[ \frac{170 ~GeV}{m_t} \right]^{0.575}
\left[ \frac{Br(K_L)}{3\cdot 10^{-11}} \right]^{1/4}
\end{equation}
where
\begin{equation}\label{7}
r_{\pm}(z)=\frac{1}{z}(1\pm\sqrt{1-z^2})
\qquad
z=a,b
\end{equation}
We note that the weak dependence of $\mid V_{cb}\mid$ on $Br(\klpnn)$
allows to achieve high accuracy for this CKM element even when $Br(\klpnn)$
is not measured precisely.

As illustrative examples we consider in table 1 three scenarios.
The first four rows give the assumed input parameters and their
experimental errors. The remaining rows give the results for
selected parameters. Further results can be found in \cite{AB94}.
The accuracy in the scenario I should be achieved at B-factories,
HERA-B,
at FNAL and at KEK.
 Scenarios II and
III correspond potentially to B-physics at Fermilab during
the Main Injector era and at LHC respectively.
The experimental errors on $Br(\klpnn)$
 to be achieved in the next 15 years are most probably unrealistic,
but I show this exercise anyway in order to motivate this very
challenging enterprise.
\begin{table}
\begin{center}
\begin{tabular}{|c||c||c|c|c|}\hline
& Central &$I$&$II$&$III$\\ \hline
$\sin(2\alpha)$ & $0.40$ &$\pm 0.08$ &$\pm 0.04$ & $\pm 0.02 $\\ \hline
$\sin(2\beta)$ & $0.70$ &$\pm 0.06$ &$\pm 0.02$ & $\pm 0.01 $\\ \hline
$m_t$ & $170$ &$\pm 5$ &$\pm 3$ & $\pm 3 $\\ \hline
$10^{11} Br(K_L)$ & $3$ &$\pm 0.30$ &$\pm 0.15$ & $\pm 0.15 $\\ \hline\hline
$\varrho$ &$0.072$ &$\pm 0.040$&$\pm 0.016$ &$\pm 0.008$\\ \hline
$\eta$ &$0.389$ &$\pm 0.044$ &$\pm 0.016$&$\pm 0.008$ \\ \hline
$\mid V_{ub}/V_{cb}\mid$ &$0.087$ &$\pm 0.010$ &$\pm 0.003$&$\pm 0.002$
 \\ \hline
$\mid V_{cb}\mid/10^{-3}$ &$39.2$ &$\pm 3.9$ &$\pm 1.7$&$\pm 1.3$\\ \hline
$\mid V_{td}\mid/10^{-3}$ &$8.7$ &$\pm 0.9$ &$\pm 0.4$ &$\pm 0.3$ \\
 \hline \hline
$\mid V_{cb}\mid/10^{-3}$ &$41.2$ &$\pm 4.3$ &$\pm 3.0$&$\pm 2.8$\\ \hline
$\mid V_{td}\mid/10^{-3}$ &$9.1$ &$\pm 0.9$ &$\pm 0.6$ &$\pm 0.6$ \\
 \hline
\end{tabular}
\end{center}
\centerline{}
\caption{Determinations of various parameters in scenarios I-III }
\end{table}
Table 2 shows very clearly the potential of CP asymmetries
in B-decays and of $\klpnn$ in the determination of CKM parameters.
It should be stressed that this high accuracy is not only achieved
because of our assumptions about future experimental errors in the
scenarios considered, but also because $\sin(2\alpha)$ is a
very sensitive function of $\varrho$ and $\eta$ \cite{BLO},
$Br(\klpnn)$
depends strongly on $\mid V_{cb}\mid$ and most importantly because
of the clean character of the quantities considered.

It is instructive to investigate whether the use of
$\kpnn$ instead of $\klpnn$ would also give interesting results for
$V_{cb}$ and $V_{td}$. Afterall $\kpnn$ will certainly be seen
before $\klpnn$.
 We again consider scenarios I-III with
$Br(\kpnn)= (1.0\pm 0.1)\cdot 10^{-10}$ for the scenario I and
$Br(\kpnn)= (1.0\pm 0.05)\cdot 10^{-10}$ for scenarios II and III
in place of $Br(\klpnn)$ with all other input parameters unchanged.
An analytic formula for $\vcb$ can be found in \cite{AB94}.
The results for $\varrho$, $\eta$, and $\mid V_{ub}/V_{cb}\mid$
remain of course unchanged. In the last two rows of table 2 we show the
results for $\mid V_{cb} \mid$ and $\mid V_{td}\mid$ .
We observe that
due to the uncertainties present in the charm contribution to
$\kpnn$, which was absent in $\klpnn$, the determinations of
 $\mid V_{cb}\mid$ and  $\mid V_{td}\mid$  are less accurate.
 If the uncertainties due to the charm mass
and $\Lambda_{\overline{MS}}$ are removed one day this analysis
will be improved \cite{AB94}.

An alternative strategy is to use the measured value of $R_t$ instead
of $\sin(2\alpha)$. Then (\ref{5})
is replaced by
\begin{equation}\label{5a}
\bar\varrho = 1-\bar\eta r_{+}(b)\quad ,\quad
\bar\eta=\frac{R_t}{\sqrt{2}}\sqrt{b r_{-}(b)}
\end{equation}
The result of this exercise is shown in table 3.
Again the last two rows give the results when $\klpnn$ is replaced
by $\kpnn$.
Although this determination of
CKM parameters cannot fully compete with the previous one the
consistency of both determinations will offer an important test of
the standard model.

\begin{table}
\begin{center}
\begin{tabular}{|c||c||c|c|c|}\hline
& Central &$I$&$II$&$III$\\ \hline
$R_t$ & $1.00$ &$\pm 0.10$ &$\pm 0.05$ & $\pm 0.03 $\\ \hline
$\sin(2\beta)$ & $0.70$ &$\pm 0.06$ &$\pm 0.02$ & $\pm 0.01 $\\ \hline
$m_t$ & $170$ &$\pm 5$ &$\pm 3$ & $\pm 3 $\\ \hline
$10^{11} Br(K_L)$ & $3$ &$\pm 0.30$ &$\pm 0.15$ & $\pm 0.15 $\\ \hline\hline
$\varrho$ &$0.076$ &$\pm 0.111$&$\pm 0.053$ &$\pm 0.031$\\ \hline
$\eta$ &$0.388$ &$\pm 0.079$ &$\pm 0.033$&$\pm 0.019$ \\ \hline
$\mid V_{ub}/V_{cb}\mid$ &$0.087$ &$\pm 0.014$ &$\pm 0.005$&$\pm 0.003$
 \\ \hline
$\mid V_{cb}\mid/10^{-3}$ &$39.3$ &$\pm 5.7$ &$\pm 2.6$&$\pm 1.8$\\ \hline
$\mid V_{td}\mid/10^{-3}$ &$8.7$ &$\pm 1.2$ &$\pm 0.6$ &$\pm 0.4$ \\
 \hline \hline
$\mid V_{cb}\mid/10^{-3}$ &$41.3$ &$\pm 5.8$ &$\pm 3.7$&$\pm 3.3$\\ \hline
$\mid V_{td}\mid/10^{-3}$ &$9.1$ &$\pm 1.3$ &$\pm 0.8$ &$\pm 0.7$ \\
 \hline
\end{tabular}
\end{center}
\centerline{}
\caption{ As in table 2 but with $\sin(2\alpha)$ replaced by $R_t$.}
\end{table}

Of particular interest will be the comparison of $\mid V_{cb}\mid$
determined as suggested here with the value of this CKM element extracted
from tree level semi-leptonic  B-decays.
 Since in contrast to
$\klpnn$ and $\kpnn$, the tree-level decays are to an excellent approximation
insensitive to any new physics contributions from very high energy scales,
the comparison of these two determinations of $\mid V_{cb}\mid$ would
be a good test of the standard model and of a possible physics
beyond it.

Precise determinations of all CKM parameters without hadronic uncertainties
 along the lines presented
here can only be realized if the measurements of CP asymmetries in
B-decays and the measurements of $Br(\klpnn)$, $Br(\kpnn)$ and $x_d/x_s$
can reach the desired accuracy.
All efforts should be made to achieve this goal.

\section{Final Remarks}
In this review we have discussed the most interesting quantities which
when measured should have important impact on our understanding of the CP
violation and of the quark mixing. We have discussed both CP violating and
CP conserving loop induced decays because in the standard model CP violation
and quark mixing are closely related.

In this  review we have concentrated on rare decays and  CP violation in the
standard model. The structure of rare decays and of CP violation in
extensions of the
standard model may deviate from this picture.
Consequently the situation in this field could turn out to be very different
from the one presented here.
However in order to distinguish the standard model predictions from
the predictions of its extensions it is essential that the
theoretical calculations reach acceptable precision. In this
context we have emphasized the importance of the QCD calculations in
rare and CP violating decays. During the recent years a considerable
progress has been made in this field through the computation of NLO
contributions to a large class of decays. This effort reduced considerably
the theoretical uncertainties in the relevant formulae and thereby improved
the determination of the CKM parameters to be achieved in future
experiments. At the same time it should be stressed that whereas the
theoretical status of QCD calculations for rare semileptonic decays like
$K \to \pi\nu\bar\nu$, $B\to \mu\bar\mu$, $B \to X_s e^+ e^-$
 is fully satisfactory and
the status of $B\to X_s\gamma$ should improve in the coming years, a lot
remains to be done in a large class of non-leptonic decays or transitions
where non-perturbative uncertainties remain sizable.

\section{Acknowledgements}

I would like to thank the organizers for inviting me to this symposium
and for their great hospitality.
The splendid birthday party " Chez Zalewskis " will never be forgotten.
Finally I would like to thank all the members of the Munich-NLO-Club
for the great time we had and still have together.

\vfill\eject

\end{document}